\documentclass{ijmpi_authors}

\usepackage{dcolumn}% Align table columns on decimal point

\newcommand{\spion}{\mathrm{NP}}
\newcommand{\hdrive}{H_\mathrm{drive}}
\newcommand{\idrive}{I_\mathrm{drive}}
\newcommand{\bspion}{B_\mathrm{\spion}}
\newcommand{\bspionvec}{\vec{B}_\mathrm{\spion}}
\newcommand{\omegarf}{f_\mathrm{rf}}
\newcommand{\omegal}{f_\mathrm{prec}}
\newcommand{\omegaspion}{f_\mathrm{\spion}}
\newcommand{\mfe}{m_\mathrm{Fe}}
\newcommand{\mfetot}{m_\mathrm{Fe}^\mathrm{tot}}
\newcommand{\mfespec}{m_\mathrm{Fe}^\mathrm{spec}}
\newcommand{\Mspion}{M_\mathrm{\spion}} % Magnetisation
\newcommand{\muspion}{\mu_\mathrm{core}}

\newcommand{\musample}{\mu_\mathrm{sample}}
\newcommand{\musamplesat}{\mu_\mathrm{sample}^\mathrm{sat}}
\newcommand{\mcore}{m_\mathrm{core}}
\newcommand{\Nspion}{N_\mathrm{\spion}}
\newcommand{\vspion}{V_\mathrm{core}}

\newcommand{\rhocore}{\rho_\mathrm{core}}

\newcommand{\opm}{\mathrm{OPM}}
\newcommand{\emg}{\mathrm{EMG}{-}707}
\begin{document}

\twocolumn[
\title{M(H) dependence and size distribution of SPIONs measured by atomic magnetometry}

\author{Colombo}{Simone}{a,\ast}
\author{Lebedev}{Victor}{a}
\author{Gruji\'c}{Zoran~D.}{a}
\author{Dolgovskiy}{Vladimir}{a}
\author{Weis}{Antoine}{a}

\affiliation{a}{D\'{e}partement de Physique, Universit\'{e} de Fribourg, Chemin du Mus\'{e}e 3, 1700 Fribourg, Switzerland}
\affiliation{\ast}{Corresponding author, email: simone.colombo@unifr.ch}

\maketitle

\begin{abstract}
We demonstrate that the quasistatic recording of the magnetic excitation function $M(H)$ of superparamagnetic iron oxide magnetic nanoparticle (SPION) suspensions by an atomic magnetometer allows a precise determination of the sample's iron mass content $\mfe$ and the particle size distribution.
\end{abstract}
]
% =======================================================================

% You should use BibTeX and apsrev.bst for references
% Choosing a journal automatically selects the correct APS
% BibTeX style file (bst file), so only uncomment the line
% below if necessary.
%\bibliographystyle{apsrev}

%\begin{document}

%\title{Direct measurement of magnetic nanoparticle $M(H)$ dependencies using an atomic magnetometer.}
%\date{\today}

%


%\textbf{I suggest Appl. Phys. Lett. as journal for this
%communication. Note, however, that APL papers do not have sections
%and subsections}

%\section{Introduction}\label{sec:intro}
\section{Introduction}
Magnetic nanoparticles (MNP) play a role of increasing importance in biomedical and biochemical applications \cite{HuangJNR2011}.
The use of MNPs in hyperthermia \cite{DeatschJMMM2014} and as MRI contrast agents \cite{LeeCSR2012} is well established, and active studies continue in view of using MNPs for targeted drug delivery \cite{ArrueboNT2007,NeubergerJMMM2005,MahmoudiADDR2011}.
Most MNP applications call for a quantitative characterization and monitoring of the particle distributions both prior to and after their administration into the biological tissue.
Two imaging modalities for determining MNP distributions in biological tissues are being actively pursued, viz., magnetorelaxation (MRX)~\cite{squidPTB2014} and Magnetic Particle Imaging (MPI) \cite{panagiotopoulosIJN2015}.

The superparamagnetic character of the MNPs' magnetic $M(H)$ response makes magnetic measurements the method of choice for their investigation.
High-sensitivity magnetic induction detection plays a key role in view of minimizing the administered MNP dose in biomedical applications.
Established MNP characterization/detection methods mainly rely on detecting the oscillating induction $B(t){\propto}M(t)$ induced by a harmonic excitation $H(t)$ with a magnetic pick-up (induction) coil.

Here we describe our successful attempt to replace the pick-up coil by an atomic magnetometer  which allows recording quasi-static $B(t)$ variations in frequency ranges that are not accessible to induction coils.
Since their introduction in the 1950s \cite{DehmeltPR1957}, atomic magnetometers, also known as optical or optically-pumped magnetometers ($\opm$) have become important instruments with a broad range of applications  \cite{Budker}.
 %review applications ranging in fundamental metrology \cite{GreenNIMPR1996} and geoprospection \cite{UnterbergerJGR1960} to  \cite{XuPNAS2006}. More recently they have found successful use in biomedical studies (\cite{BisonAPL2009,SanderBOE2012,AlemPMB2015}).
%
%The $\opm$s' comparatively simple design (portable and miniaturized variants without major performance compromises have been demonstrated \cite{Budker}), together with their intrinsic sensitivity in the femtoTesla range,
%%and undemanding usage conditions allowing marine, airborn and spaceborn applications,
%permit their easy integration into existing instruments as alternative or complementary sensors \cite{OidaJMR2012}.
%
%Strong magnetism of of those particles facilitates magnetic methods of their detection.
%%
%However superparamagnetic character of particles and low amounts to be detected needs special sensitive and MNP-selective methods to be deployed.
%
%
Reports on applications of $\opm$s for studying MNPs are scarce and have, so far, focused on MRX studies \cite{MaserRSI2011,JohnsonJMMM2012,DolgovskiyJMMM2015}.
%

%While many variants of $\opm$ has been proposed and successfully realised in past decades \cite{Budker}, high bandwidth $\opm$s [Ref ] are less commonly deployed, and to our knoledge have never been used for either relaxometric or susceptomretic MNP detection.

%
%introduction......why what how who else
%
%.... MORE ....

We have studied the magnetic response $M(H)$ of water-suspended
superparamagnetic iron oxide nano-particle samples exposed to
time varying excitation fields $\hdrive(t)$.
We show that an  $\opm$ can be used to record the
magnetic induction $\bspion(t)$ produced by (and proportional to)
the time-varying MNP magnetization~$\Mspion(t)$, itself proportional to the iron mass content of the sample.
%
%%%%%%%%%%%%%%%%%%%%%%%%%%%%%%%%%%%%%%%%%%%%%%%%%%%%%
\section{Apparatus}
\begin{figure}[h]
  \centering
  \includegraphics[width=1\columnwidth]{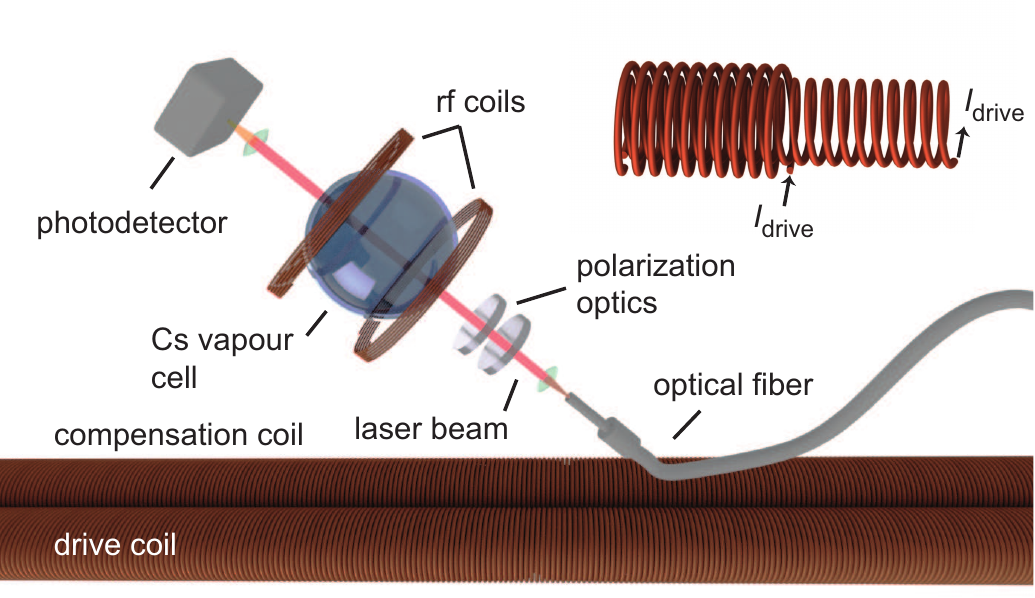}%SetupV.pdf
  \caption{Experimental set-up of the $\opm$-based MPS.
  Top right inset: Current flow in the two (opposite handedness) layers in one of the two identical double-layered solenoids.
  The currents flow in opposite directions in the second double solenoid.}
  \label{fig.setup}
\end{figure}

%%%%%%%%%%%%%%%%%%%%%%%%%%%%%%%%%%%%%%%%%%%%%%%%%%%%%
The experiments were carried out using the apparatus sketched in
Fig.~\ref{fig.setup} that was mounted  in a double aluminum
chamber of walk-in size, described by Bison \emph{et al.}~\cite{BisonAPL2009}.
%
%
%\section{Experimental setup}\label{sec:setup}
%
A major challenge for operating an $\opm$-based magnetic
particle spectrometer lies in the fact that the $\opm$ has to record
fields $\bspion$ in the pT$\dots$nT range, while being placed as closely
as possible to the drive coil producing fields $\hdrive$ of
several mT/$\mu_0$.

A 70~cm long solenoid with an aspect ratio of 50:1
produces an oscillating drive field $\hdrive(t)$ with an amplitude of up to ${\sim}$16 mT$_\mathrm{pp}/\mu_0$.
The drive solenoid was wound as a double layer of 1.2 mm diameter copper wire on a PVC tube.
The two layers have opposite handedness, such that the longitudinal currents originating from the coil's helicoidal structure
(the individual wire loops are not perfectly perpendicular to the solenoid axis) cancel.
A second, identical, but oppositely{-}poled double solenoid, placed next to the drive solenoid, strongly suppresses the stray field originating from the solenoid's finite aspect ratio.
These passive measures reduce the total stray field at the OPM position (at a distance $R\sim$7~cm from the solenoids)
by a factor of 10$^6$ compared to the field inside of the excitation solenoid.

The $\opm$ module is similar to the one described by Bison \emph{et al.}~\cite{BisonAPL2009}, except that the rf field is oriented along the light propagation direction.
The sensor uses room-temperature Cs vapour contained in
a $\sim$30~mm diameter evacuated and paraffin-coated glass cell.
The $\opm$ is operated as a so-called $M_x$ magnetometer, in
which a weak  magnetic field (rf field) oscillating at frequency
$\omegarf$ (produced by a pair of small Helmholtz coils) drives the
precession of the Cs vapour's spin polarization around a bias
magnetic field $\vec{B}_0$ in a resonant, phase-coherent
manner.
A single circularly-polarized laser beam ($\lambda${=}894~nm),
locked to the 4{-}3 hyperfine component of the D$_1$ transition
serves both to create the spin polarization and to detect its
precession by monitoring the synchronous modulation of the
transmitted laser power.
An electronic phase-locked loop (PLL) consisting of a phase detector
and a voltage-controlled oscillator ensure that $\omegarf$ stays
phase-locked to the spin precession frequency
%
%\begin{eqnarray}
%\omegal
%%=\gamma_F\,|\vec{B}_\mathrm{tot}|
%&{=}& \gamma_F\,|\vec{B}_0{+}\delta\bspionvec|%\nonumber\\
%%
%\approx\gamma_F\left(|\vec{B}_0\,|{+}\,\delta\bspionvec{\cdot}\hat{B}_0\right)\nonumber\\
%%
%&{\equiv}&f_0+\delta\omegaspion\,,
%\label{eq:gyromag}
%\end{eqnarray}
%
\begin{equation}
\omegal
%=\gamma_F\,|\vec{B}_\mathrm{tot}|
{=}\gamma_F\,|\vec{B}_0{+}\delta\bspionvec|%\nonumber\\
\approx\gamma_F\left(|\vec{B}_0\,|{+}\,\delta\bspionvec{\cdot}\hat{B}_0\right)%\nonumber\\
{\equiv}f_0+\delta\omegaspion\,,
\label{eq:gyromag}
\end{equation}
where $\gamma_F\approx$ 3.5~Hz/nT, so that
$f_0{=}$95~kHz in the used bias field $B_0$ of 27~$\mu$T.
%Eq.~\ref{eq:gyromag} translates to .
%
The phase detection and PLL are implemented using a digital lock-in amplifier (Zurich Instruments, model HF2LI, DC---50~MHz) which provides a direct numerical output of the deviations $\delta\omegaspion$ that are proportional to the signal of interest $\delta\bspion$.
Although the $M_x$ magnetometer is scalar in its nature, it is---to
first order---sensitive only to the projection of
$\delta\bspionvec$ onto $\vec{B}_0$, since
$|\delta\bspionvec|{\ll}|\vec{B}_0|$.
It thus acts as a vector component magnetometer like a SQUID.

The bias field $\vec{B}_0$ is oriented parallel to the solenoids, in
order to maximize the sensitivity to the induction $\delta\bspionvec$ of
interest.
The effect of the solenoids' stray field on $\vec{B}_0$
results in a harmonic oscillation of $|\vec{B}_0|$ with an amplitude of ${\approx}$3~nT.
%
%This extra field is later suppressed by a reference measurement.
%
The magnetometer detects no signal without MNP sample down to its noise floor of~$\approx$5~pT$/\sqrt{Hz}$.
Under typical experimental conditions the magnetometer can react to magnetic field changes with a
bandwidth of $\sim$1~kHz, while keeping the mentioned sensitivity.

%The solenoids' residual stray field component along $\hat{B}_0$ is
%further suppressed by a feedforward scheme, in which a
%rheostat-controlled fraction of the time-dependent current producing
%$\hdrive$ is added, after an appropriate phase shift, to the current
%producing the static bias field $\vec{B}_0$.
%%
%
%%%%%%%%%%%%%%%%%%%%%%%%%%%%%%%%%%%%%%%%%%%%%%%%%%%%%
\section{Measurements and results}
\begin{figure}[h]
  \centering
  \includegraphics[width=.9\columnwidth]{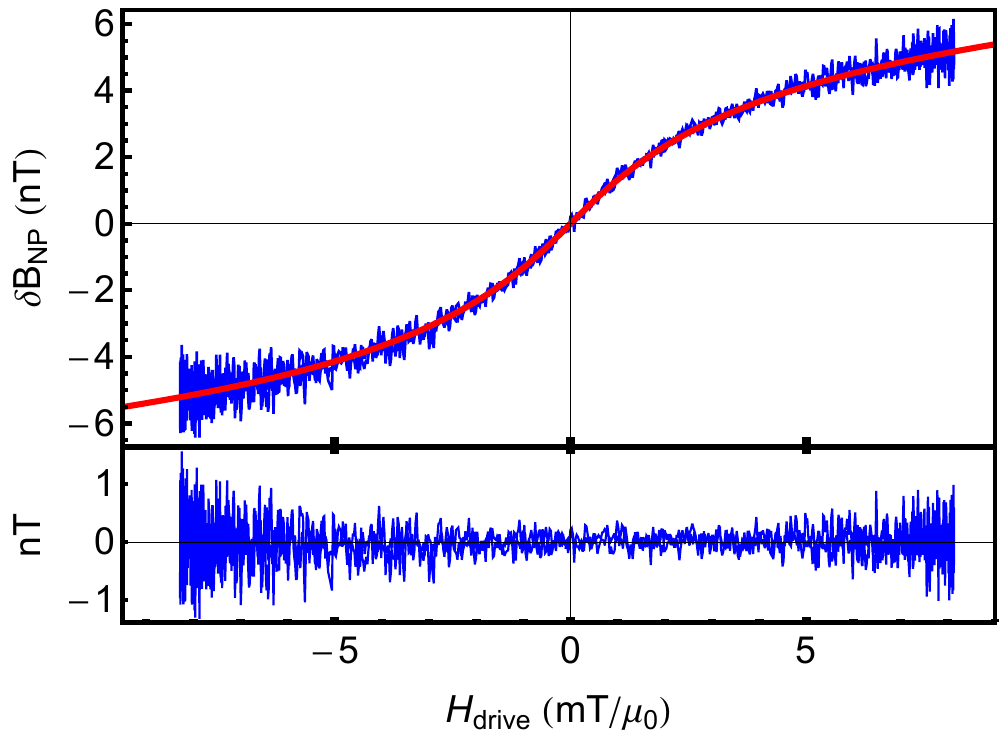}
    \caption{Typical $\delta\bspion(\hdrive)$ response of a 0.5~ml Ferrotec EMG${-}707$ sample containing 0.3~mg of iron.
  The data, representing 50 averaged sinusoidal $\hdrive(t)$ cycles (total recording time of $\sim$5~s) are fitted by Eq.~\ref{eq:bmnp1}.
  The lower graph shows the fit residuals.}
  \label{fig:Langevin}
\end{figure}

%%%%%%%%%%%%%%%%%%%%%%%%%%%%%%%%%%%%%%%%%%%%%%%%%%%%%
%
We have performed measurements on different MNP samples suspended in aqueous solutions of $\sim$500~$\mu$l contained in sealed glass vessels.
The samples can be moved freely through one of the solenoids, and positioned in its center by maximizing the magnetometer signal.
Experiments were done by driving the solenoids with a sine-wave-modulated current provided by a high current operational amplifier (Texas Instruments, model OPA541, 5~A max.).
We recorded time series (2000 samples per period) of both the magnetometer signal, i.e., its oscillation frequency change $\delta\omegaspion$ and the coil current $\idrive$, proportional to the drive field $\hdrive(t)$, monitored as the voltage drop over a series resistor.
An x--y representation  of $\delta\omegaspion$ vs.~$\idrive$ that is equivalent to $\delta\bspion$ vs.~$\hdrive$ after calibration, can then directly be visualized as an oscilloscope trace.
Time series of $\delta\omegaspion(t)$ and $\idrive(t)$ are stored for further off-line processing.
Figure \ref{fig:Langevin} shows a typical example of a recorded $\delta\bspion(\hdrive){\propto}\Mspion(\hdrive)$ dependence, together with a fitted function and the fit residuals.

%
%Typical $M(H)$ curve of the studied samples shows high signal-to-noise response, and can be accurately fit by Langevin function:
%
%\begin{eqnarray}
%\delta\bspion(\hdrive) = c \left( \coth \frac{\hdrive}{H_K}-\frac{H_K}{\hdrive} \right)\,,
% \label{eq:Langevin}
%\end{eqnarray}
%%
%where  $H_K$ is the sample-specific constant.

Each MNP has a magnetic moment $\muspion{=}\vspion\,M_S$, where  $\vspion{=}4\pi r^3/3$ is the particle's core volume and $M_s$ its saturation magnetization.
The infinitesimal contribution of particles with radius $r$ to the total magnetic moment is given by
\begin{eqnarray}
\mathrm{d}\musample(H;r)~{=}&\mathrm{d}\muspion(r)\mathcal{L}\left(\frac{H}{H_k(r)}\right)\nonumber\\
{=}&\mathrm{d}\Nspion(r)\vspion(r) M_S\mathcal{L}\left(\frac{H}{H_k(r)}\right)\,,
\label{eq:musample1}
\end{eqnarray}
where the Langevin function $\mathcal{L}(x)=\coth(x)-x^{-1}$ describes the field-dependent degree of magnetization.
The saturation field
\begin{equation}
H_k(r)=\frac{k_BT}{\mu_0\muspion}=\frac{3k_BT}{4\pi\mu_0r^3\,M_s}
\end{equation}
is a property of the individual particle.
The scaling prefactor in Eq.~\ref{eq:musample1}, expressed as
$\mathrm{d}\Nspion\,\vspion{=}\mathrm{d}\mfetot/(\alpha\,\rhocore)$,
shows that the contribution $\mathrm{d}\mu_{\mathrm{sample}}$ depends in a linear manner on its total iron mass content $\mathrm{d}\mfetot$.
In the last expression $\alpha{=}\mfe/\mcore\approx$0.71 is the mass fraction of iron in each nanoparticle's core,  $\rhocore{\equiv}\mcore/\vspion$ and  $\mcore$ the core's density and mass, respectively.
With this, Eq.~\ref{eq:musample1} takes the form
\begin{equation}
\mathrm{d}\musample(H;r){=}\frac{\mathrm{d}\mfetot\,M_S}{\alpha\,\rhocore}\mathcal{L}\left(\frac{H}{H_k(r)}\right)\,.
\label{eq:musample2}
\end{equation}
In practice, the MNP sample shows a size polydispersity that we describe by the lognormal distribution
\begin{equation}
w_{\mathrm{LN}}(r; \mu, k)=\frac{1}{r \sqrt{2\pi k}}\exp\left(-\frac{\ln^2(r/\mu)}{2\,k}\right)\,.
\label{eq:SZ}
\end{equation}
The infinitesimal mass of iron in particles of a given size is then expressed as
\begin{eqnarray}
\mathrm{d}\mfetot~=~\mfetot~\frac{r^3 w_{\mathrm{LN}}(r;\mu,k)}{\mu^2e^{\frac{9}{2}k}}\mathrm{d}r=~\mfetot~D(r;\mu,k)~\mathrm{d}r
\end{eqnarray}
where $D(r;\mu,k)$ is the mass fraction distribution with mean radius $\overline{r}=\mu~e^{7/2k}$ and standard deviation $\sigma{=}\overline{r}\sqrt{e^k-1}$.
The distribution $D$ is readily extended to multimodal variants by
\begin{equation}
D_{\spion}^{(n)}(r)=\,\sum_{i=1}^n A_i D(r;\mu_i,k_i)
\label{eq:bimodal}
\end{equation}
where $A_i$ is the relative mass ratio of the mode $i$ with $\sum_{i=1}^nA_i{=}1$ and $n$ is the number of modes of the distribution.
The magnetic moment of a polydisperse sample is then given by
$
\musample(H){=}\int_0^\infty\mathrm{d}\musample(H;r)\,.
$
The $\opm$ detects the far-field magnetic induction
\begin{eqnarray}
\delta\bspion(H)&=&\frac{\mu_0}{4\pi}\frac{\musample(H)}{R^3}=\frac{\mu_0}{4\pi R^3}\musamplesat\,\nonumber\\
&\times&\int\limits_{0}^{\infty}D^{(n)}_{\spion}(r)\mathcal{L}\left(\frac{4\pi\mu_0 r^3M_SH}{3k_BT} \right)\mathrm{d}r\,,
\label{eq:bmnp1}
\end{eqnarray}
were the saturation ($H{\gg}H_k$) value of the sample's magnetic moment is given by
\begin{equation}
 \musamplesat{=}\frac{M_S}{\alpha\rhocore}\,\mfetot\,.
\label{eq:musamplesatDef}
\end{equation}
%
%The $M(H)$ dependence of a monodisperse (particle radius $r$) distribution of MNPs suspended in a solvent of volume $\vsample$ is described by the Langevin function
%%
\begin{figure}[b]
  \centering
  \includegraphics[width=.9\columnwidth]{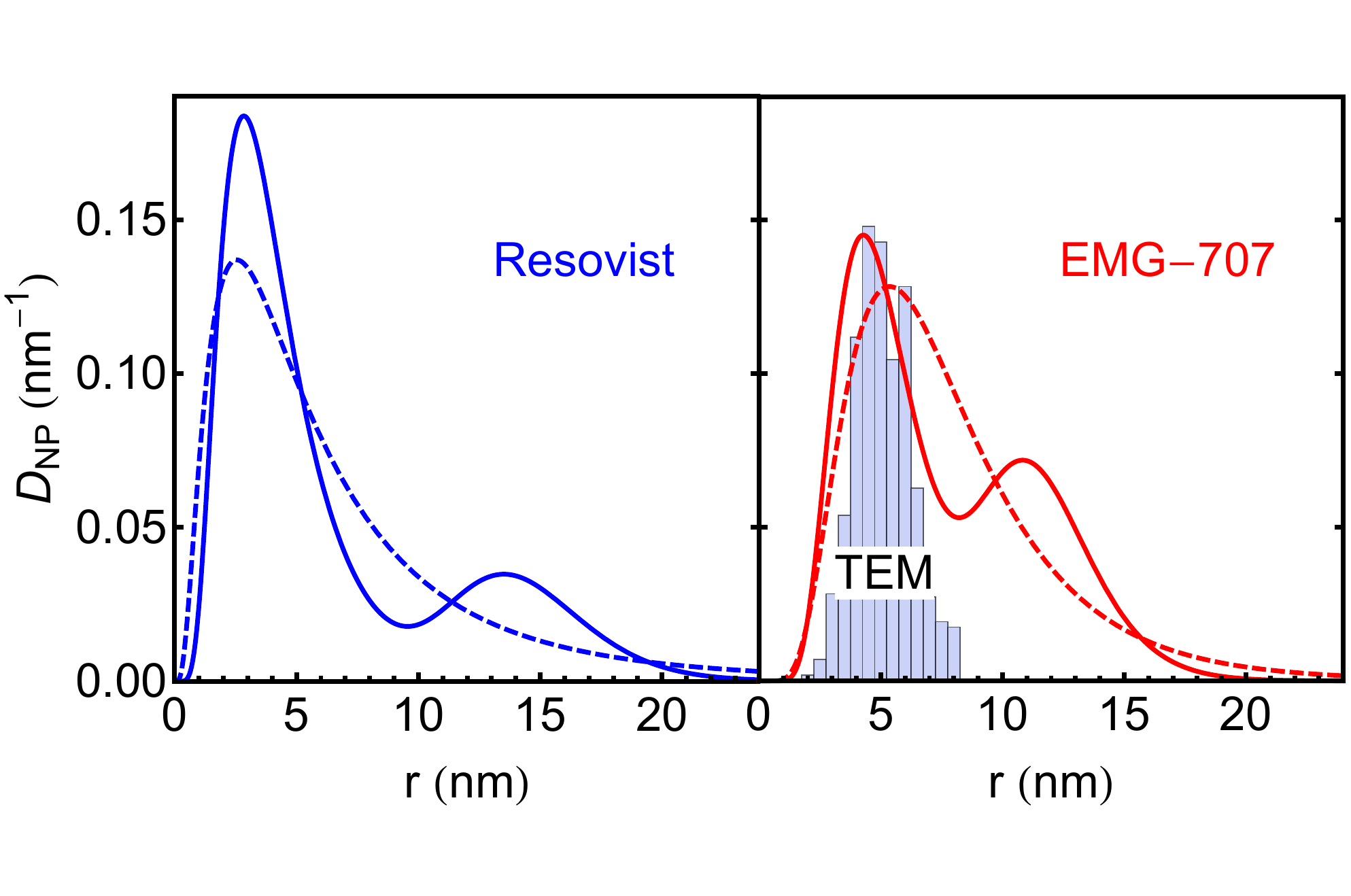}%{SZResults.eps}
  \caption{Mono- and bi-modal MNP mass fraction distributions inferred from fits of Eq.~\ref{eq:bmnp1} to recorded $M(H)$ curves.}
  \label{fig:sizeDist}
\end{figure}
We have performed $M(H)$ measurements like the one shown in Fig. \ref{fig:Langevin} on 500~$\mu$l samples in a dilution series of the ferrofluids EMG${-}707$ (from Ferrotec)  and Resovist.
The experimental parameters ($R$, $T$) and the sample parameters ($\rhocore$, $\alpha$) are known \emph{a priori}.
We use the samples with the highest iron content to obtain the MNP parameters ($M_S$ and mass fraction distribution) in the following way:
We fit Eq.~\ref{eq:bmnp1} to the data by fixing the amount of iron as a known parameter derived from manufacturer specifications and the degree of dilution, i.e., $\mfetot{=}\mfespec{=}$4.2~mg for Resovist and $\mfetot{=}\mfespec{=}$10.2~mg for  EMG${-}707$, keeping the saturation magnetization $M_S$ and the particle mass fraction distribution parameters $\bar{r}_i$  and $\sigma_i$ as fit parameters.
We have performed these calibrations assuming both mono-modal ($n{=}1$) and bi-modal ($n{=}2$) mass fraction distributions.
The results are listed in Table~\ref{tab:distfits}.

Assuming a monomodal distribution, the fits yield $M_S$ and the mass fraction distribution parameters.
However, this procedure leads to $M_S$ values that are much smaller than the literature/manufacturer values (rows R-1 and E-1 in Table 1).
The reason for this discrepancy lies in the fact that because of the modest $\hdrive$ field amplitudes used in our experiment we do not strongly saturate the $\mathcal{L}(\hdrive)$ dependence, so that $M_S$ is determined by the linear slope of the $\mathcal{L}(\hdrive)$ dependence.
In the $\hdrive{\approx}0$ region, $M_S$ is strongly correlated with the mass fraction distribution parameters.

\begin{table}
	\centering
	\caption{Results from fitting magnetization of Resovist (R) and EMG{-}707 (E) with mono(1)- and bi(2)-modal distributions.
Fixed parameters are marked by (f).
The distributions of the fits marked with $^{\ast}$ are shown in Figure \ref{fig:sizeDist}.
 }
	\label{tab:distfits}	
	\begin{tabular}{l c c c c c}
%    %\hline
			         & $M_S$          & $\bar{r}_1/\sigma_1$& $\bar{r}_2/\sigma_2$& $A_1$ & FOM  \\
			         & kA/m           & nm           	     & nm                  &       & nT   \\
 %   \hline
        R-1          & 143            & 15.4/5.8       & --/--        & --	   & 0.45      \\
		R-1$\ast$    & 340(f)         &7.0/6.8       & --/--        & --    & 0.78\\
        R-2$\ast$    & 340(f)         & 4.3/2.4    & 14.5/2.8       & 0.78  & 0.43\\
%  \hline
        E-1          & 279            & 11.2/3.5       & --/--        & --    & 1.01\\
        E-1$\ast$    & 418(f)         & 7.9/4.3       & --/--        & --    & 1.84\\
        E-2$\ast$    & 418(f)         & 5.3/2.1    & 11.8/2.3       & 0.64  & 0.98
	\end{tabular}
\end{table}

In a next step we have fixed (for the fits) the $M_S$ values to literature values (340~kA/m~\cite{eberbeck2013multicore} for Resovist and 418~kA/m~\cite{ferrotecAppNote} for $\emg$), leaving only  the mass fraction distribution parameters as fit parameters.
The results for monomodal distributions are listed in rows {R-1$\ast$} and {E-1$\ast$} of Table~\ref{tab:distfits}, with corresponding mass fraction distributions shown as dashed lines in Fig.~\ref{fig:sizeDist}.
However, the fit qualities of the $M(H)$ dependences obtained with  fixed $M_S$ values are worse than with  free, i.e., fitted $M_S$ values, as evidenced by the standard deviations (FOM=figure of merit, listed in Table 1) of the fit residuals.

We next have fitted bi-modal distributions.
 %
 %For this we  assume that the smallest radius mode is the one of the original solution, whose parameters can be taken from TEM size distribution (example shown for $\emg$ in Fig.~\ref{fig:sizeDist}) found in the literature (Ref.~\cite{eberbeck2013multicore} for Resovist and in Ref.~\cite{ferrotecAppNote} for $\emg$).
%
The bimodal fits yield the  $A_1$, $\bar{r}_1$, $\bar{r}_2$, $\sigma_1$ and $\sigma_2$ values listed in Table 1 as rows R-2 and E-2, respectively.
The corresponding distributions are shown as solid lines in Fig.~\ref{fig:sizeDist}.
These bimodal fits yield the best FOM of the three outlined procedures.
For Resovist the extracted parameters are in agreement with the results of Eberbeck \emph{et al.}~\cite{eberbeck2013multicore}.
For $\emg$ an agreement of the smaller size mode with the histogram given in Ref.~\cite{ferrotecAppNote} is found.
%Since the fit procedure with bi-modal distributions yields very large model parameter uncertainties, we have opted for the following procedure:
%

Freshly produced MNP solutions are basically mono-modal, but, because of cluster formation evolve during aging to a bimodal distribution, as described e.g. in Ref.~\cite{eberbeck2013multicore}.
The method demonstrated here thus allows a quantitative monitoring of this process.

\begin{figure}[hb]
  \centering
  \includegraphics[width=0.8\columnwidth]{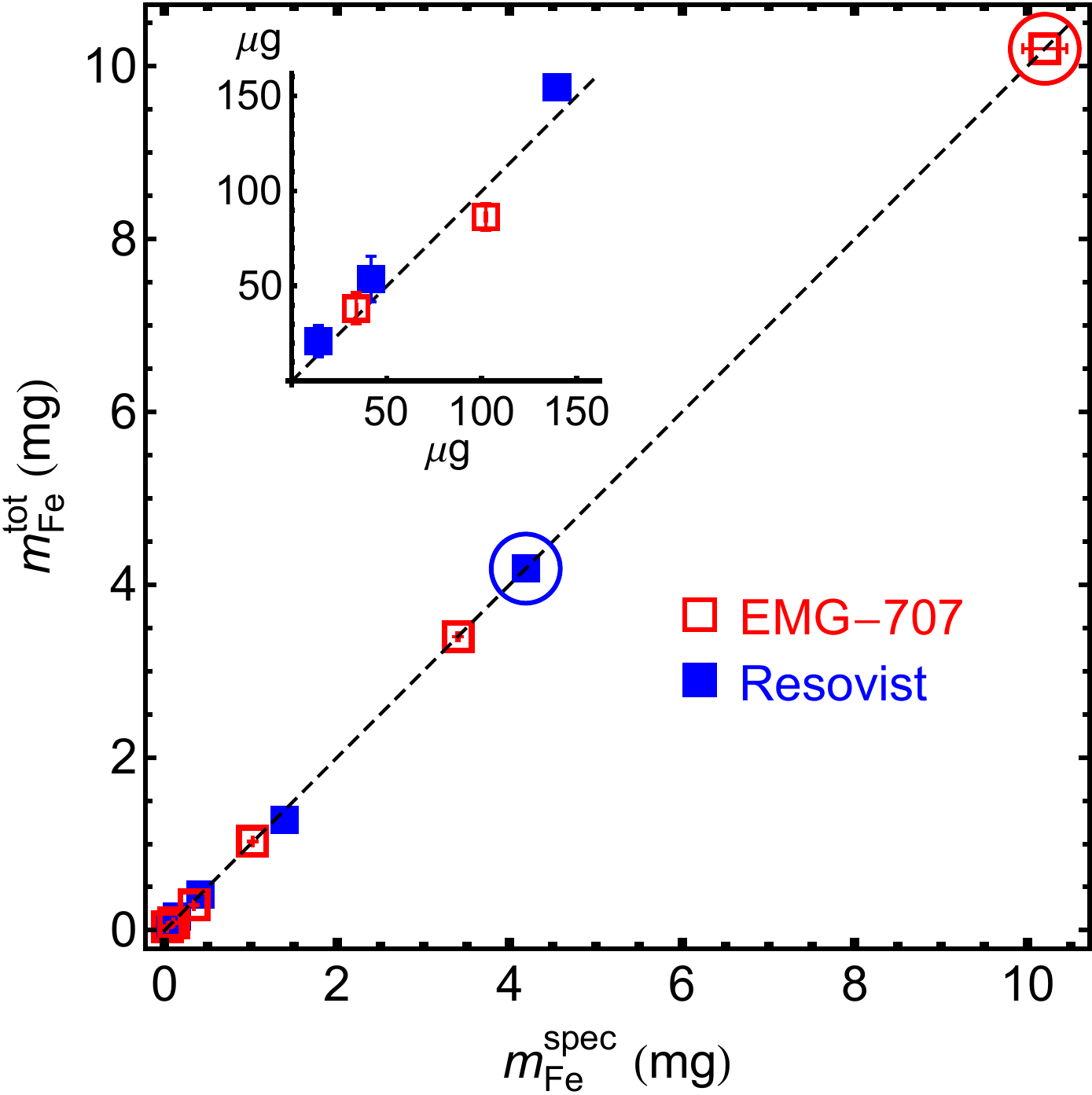}
  \caption{Dependence of the samples' iron mass $\mfetot$ (inferred from fits by Eq.~\ref{eq:bmnp1}) on the iron mass taken from manufacturer specification and degree of sample dilution, together with slope=1 linear reference line.
  Circles denote samples used for extraction of MNP parameters.
  Inset: Data points for small iron contents.
  }
  \label{fig:lCplot}
\end{figure}

With the size parameter values determined by the calibration procedure we then fit $M(H)$ curves to samples with different dilutions, having $\mfetot$ as only fit parameter.
Figure~\ref{fig:lCplot} shows the dependence of $\mfetot$ on the mass $\mfespec$ calculated from manufacturer specifications and degree of dilution.
We find an excellent agreement, as evidenced by the (non-fitted) slope{=}1 dashed line in the figure.
From the low iron content data points we estimate the current sensitivity to be on the order of $\mfetot\lesssim$7~$\mu$g.

\textbf{Acknowledgements}.
%This work was s
Supported by the Swiss National Science Foundation Grant No. 200021\_149542.

\bibliography{IWMPI2016-MH}

\end{document}